\documentclass[%
superscriptaddress,
preprint,
nofootinbib,
 amsmath,amssymb,
 aps,
]{revtex4-1}
\usepackage[section]{placeins}
\usepackage{graphicx}% Include figure files
\usepackage{dcolumn}% Align table columns on decimal point
\usepackage{bm}% bold math
\usepackage[version=3]{mhchem}
\usepackage{color}
\usepackage{float}
\usepackage{bigstrut}
\usepackage{booktabs}

\begin{document}
\sloppy
%\preprint{APS/123-QED}

\title{Isoelectronic Substitutions and Aluminium Alloying in the Ta-Nb-Hf-Zr-Ti High-Entropy Alloy Superconductor}

\author{Fabian O. von Rohr}
%\email{fabian.vonrohr@chem.uzh.ch}
\affiliation{Department of Chemistry, University of Zurich, CH-8057 Zurich, Switzerland}

\author{Robert J. Cava}
\affiliation{Department of Chemistry, Princeton University, Princeton, New Jersey 08544, USA}

\date{\today}% It is always \today, today,

%\begin{abstract} 
	%
%
%\end{abstract} 
%
\begin{abstract}
	High-entropy alloys (HEAs) are a new class of materials constructed from multiple principal elements statistically arranged on simple crystallographic lattices. Due to the large amount of disorder present, they are excellent model systems for investigating the properties of materials intermediate between crystalline and amorphous states. Here we report the effects of systematic isoelectronic replacements, using Mo-Y, Mo-Sc, and Cr-Sc mixtures, for the valence electron count 4 and 5 elements in the BCC \break Ta-Nb-Zr-Hf-Ti high entropy alloy (HEA) superconductor. We find that the superconducting transition temperature $T_{\rm c}$ strongly depends on the elemental make-up of the alloy, and not exclusively its electron count. The replacement of niobium or tantalum by an isoelectronic mixture lowers the transition temperature by more than 60 \%, while the isoelectronic replacement of hafnium, zirconium, or titanium has a limited impact on $T_{\rm c}$. We further explore the alloying of aluminium into the nearly optimal electron count \ce{[TaNb]_{0.67}(ZrHfTi)_{0.33}} HEA superconductor. The electron count dependence of the superconducting $T_{\rm c}$ for (HEA)Al$_x$ is found to be more crystalline-like than for the \ce{[TaNb]_{1-\textit{x}}(ZrHfTi)_{\textit{x}}} HEA solid solution. For an aluminum content of $x =$ 0.4 the high-entropy stabilization of the simple BCC lattice breaks down. This material crystallizes in the tetragonal $\beta$-uranium structure type and superconductivity is not observed above 1.8 K.  
\end{abstract}
 %                            % Classification Scheme.
  %                         %display desired
\maketitle
\newpage
%
%\tableofcontents
%
\section{Introduction}
High-entropy alloys (HEAs) are multi-element systems that consist of several principle constituents, crystallizing as a single phase on a simple crystallographic lattice \cite{HEA_Yeh}. A large variety of different HEAs has already been reported; they so far mainly arrange on body-centered cubic (BCC), hexagonal-closest packing (HCP), or face-centered cubic (FCC) lattices \cite{HEA_review}. These simple crystallographic lattices are stabilized by the high configurational entropy of mixing \cite{HEA_stability}. The concept of HEAs is in contrast to that of conventional alloys, which are commonly based on a single primary element that has been modestly doped in order to promote a particular desired property. There are virtually an unlimited number of potential HEA compositions. \\ \\
HEAs are currently of great research interest. They are considered for a wide variety of applications, due to their excellent specific strength \cite{HEA_hard,HEA_hard2}, superior mechanical performance at high temperatures \cite{ductile,ductile2}, and fracture toughness at cryogenic temperatures \cite{cryo,cryo2}. The concept of high-entropy stabilization has also been found to be a new potential pathway for the stabilization of chemical structures that are otherwise difficult to access. For example, the high-entropy stabilized oxide (Mg,Ni,Co,Cu,Zn)O was found to crystallize as a simple single-phase rocksalt structure \cite{HEA_oxide}. The chemical interactions in HEAs have been found to follow complex rules with long-ranged interactions causing chemical frustration \cite{frustration}. This allows the tuning of structures and properties of high-entropy alloys in a manner that is not achievable with conventional alloys and phases \cite{hcp}. \\ \\
\begin{figure}
	\centering
	\includegraphics[width=0.5\linewidth]{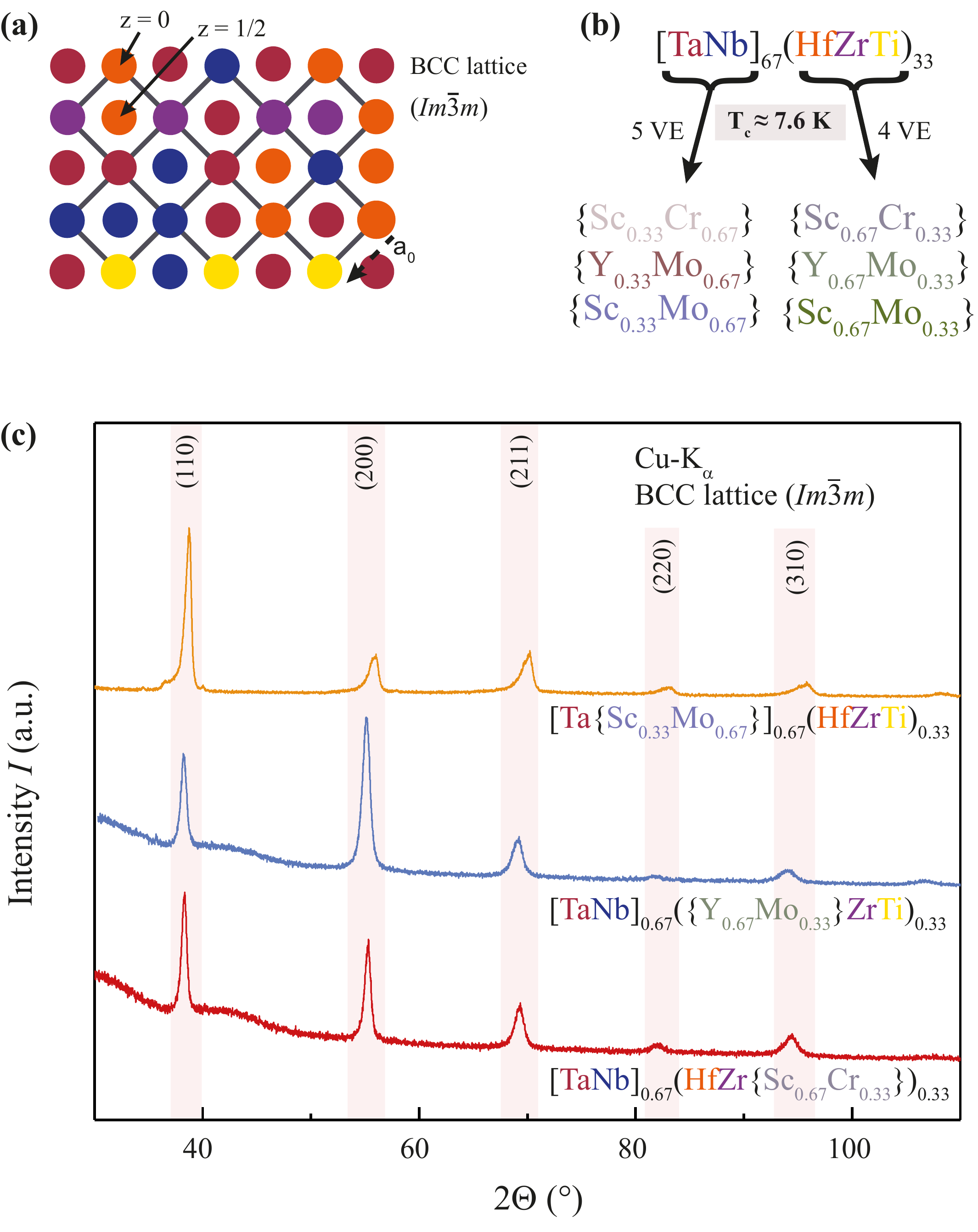}
	\caption{(a) Schematic representation of a BCC lattice with randomly distributed atoms (b) Schematic summary of the substitutions of Nb and Ta by isoelectronic {$\lbrace$\ce{Sc_{0.33}Cr_{0.67}}$\rbrace$}, {$\lbrace$\ce{Y_{0.33}Mo_{0.67}}$\rbrace$}, and {$\lbrace$\ce{Sc_{0.33}Mo_{0.67}}$\rbrace$} mixtures, and Zr, Hf, and Ti by isoelectronic {$\lbrace$\ce{Sc_{0.67}Cr_{0.33}}$\rbrace$}, {$\lbrace$\ce{Y_{0.67}Mo_{0.33}}$\rbrace$}, and {$\lbrace$\ce{Sc_{0.67}Mo_{0.33}}$\rbrace$} mixtures (c) XRD patterns for the substitutions of {$\lbrace$\ce{Sc_{0.33}Mo_{0.67}}$\rbrace$} for Nb, of {$\lbrace$\ce{Y_{0.67}Mo_{0.33}}$\rbrace$} for Hf, and of {$\lbrace$\ce{Sc_{0.67}Cr_{0.33}}$\rbrace$} for Ti.}
	\label{fig:sub_XRD}
\end{figure}  
Furthermore, HEAs have been found to be good model systems for investigating fundamental physical interactions, as their properties often fall between those of crystalline and amorphous materials, see e.g. references \cite{SG_HEA} and \cite{HEA_ferro}. Of particular interest is the recent observation of superconductivity in the HEA Ta-Nb-Hf-Zr-Ti \cite{HEA_super,HEA_theory,HEA_anneal}. This material was found to be a bulk superconductor with a critical temperature of $T_{\rm{c}} \approx$ 7.3 K. We have recently reported that the superconductivity in the HEA solid solution \ce{[TaNb]_{1-\textit{x}}(ZrHfTi)_{\textit{x}}} \cite{note} displays characteristics of both crystalline and amorphous superconductors \cite{fvrohr_PNAS}. Here, we further show that isoelectronic substitutions for the elements in this alloy, using Mo-Y, Mo-Sc, and Cr-Sc mixtures, have a profound impact on the superconducting properties, an indication of non-trivial structure-property relations. We also find that alloying the HEA Ta-Nb-Hf-Zr-Ti with aluminum, i.e. as (HEA)Al$_x$, leads to a gradual decrease of the superconducting transition temperature with decreasing electron count, and the sudden collapse of the HEA concept at $x =$ 0.4, with the formation of a complex intermetallic structure instead.
\section{Experimental}
All samples were prepared by mixing stoichiometric amounts of niobium pieces (purity 99.8 \%), tantalum foil (purity 99.9 \%), zirconium pieces (purity 99.6 \%) hafnium pieces (purity 99.6 \%), titanium slug (purity 99.95 \%), molybdenum foil (purity 99.95 \%), chromium pieces (purity 99.99 \%), scandium pieces (purity 99.9 \%), yttrium pieces (purity 99.9 \%), and aluminum shot (purity 99.99 \%). The pure metals were carefully arc melted to a single metallic nugget in an argon atmosphere with high current ($T > 2500 \ ^\circ$C). The argon atmosphere was additionally cleaned from any oxygen and water by co-heating of a zirconium sponge. The samples were melted five times for 30 s, and turned over each time, in order to ensure the optimal mixing of the constituents and a high mixing entropy. Eventually the alloys were rapidly cooled on a water-chilled copper plate. The synthesized, very hard alloys were mechanically flattened in order to measure their x-ray diffraction patterns (XRD). The pattern were obtained in a Bragg-Bretano Geometry on a Bruker D8 Advance Eco with Cu $K_\alpha$ radiation and a LynxEye-XE detector. LeBail fits for determining the average cell parameters were carried out using the FullProf program suite \cite{Fullprof}. The magnetization was studied using a \textit{Quantum Design} Physical Property Measurement System (PPMS) DynaCool equipped with a Vibrating Sample Magnetometer Option (VSM). The mechanically flattened samples were measured with the plane of the plate along the field direction in order to minimize demagnetization effects in the superconducting state.
\section{Results and Discussion}
\subsection{Isoelectronic Substitutions of the HEA Superconductor}
\begin{figure}
	\centering
	\includegraphics[width=0.5\linewidth]{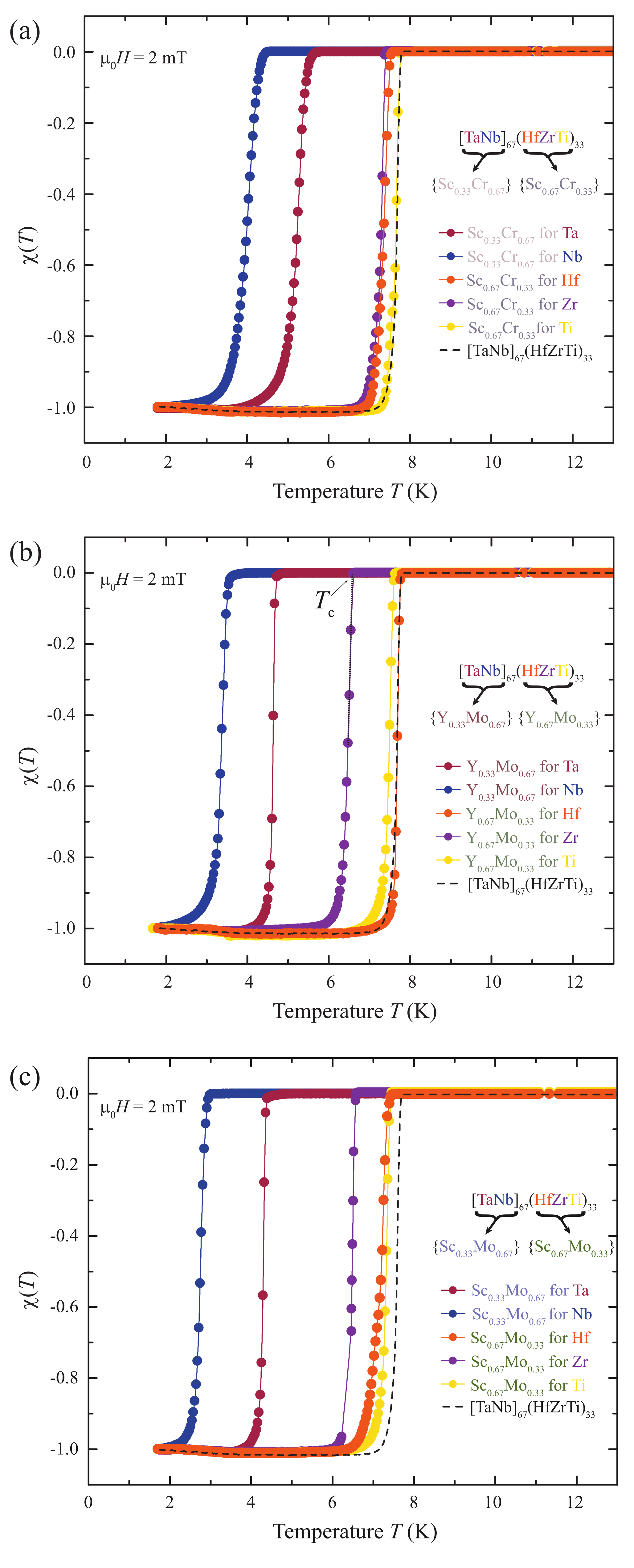}
	\caption{Temperature-dependent magnetization, in units of 1/4$\pi$ in a field of $\mu_0 H =$ 2 mT, of the pristine HEA supercondutor \ce{[TaNb]_{0.67}(HfZrTi)_{0.33}} substituted with isoelectronic (a) $\rm \lbrace Sc$-Cr$ \rbrace$, (b) $\rm \lbrace$Y-Mo$\rm \rbrace$, and (c) $\rm \lbrace$Sc-Mo$\rm \rbrace$ mixtures.}
	\label{fig:sub_mag}
\end{figure} 
\textbf{Structure of the Substituted HEA.} The effect of elemental replacements on the $T_{\rm{c}}$ of the HEA superconductor Ta-Nb-Hf-Zr-Ti was investigated by isoelectronic substitutions for the constituent atoms. For these substitutions, the electron count \cite{Felser} was kept constant, close to the previously determined optimal electron count (i.e. electrons per atom, e/a) of 4.7, which corresponds approximately to a ratio of 2:1 of the group V to group IV elements in the alloy, and a chemical formula of \ce{[TaNb]_{0.67}(HfZrTi)_{0.33}}. This electron count yields a critical superconducting transition temperature of $T_{\rm c} \approx$ 7.6 K for the pristine HEA, which crystallizes on a BCC lattice as shown in figure \ref{fig:sub_XRD}(a). The isoelectronic substitutions for the group IV (VEC 4) and group V (VEC 5) elements were carried out by using mixtures of group III (VEC 3) and group VI (VEC 6) elements. Specifically, we used {$\lbrace$\ce{Sc_{0.33}Cr_{0.67}}$\rbrace$}, {$\lbrace$\ce{Y_{0.33}Mo_{0.67}}$\rbrace$, and {$\lbrace$\ce{Sc_{0.33}Mo_{0.67}}$\rbrace$ with a VEC 5 for the substitution of niobium and tantalum, and {$\lbrace$\ce{Sc_{0.67}Cr_{0.33}}$\rbrace$, {$\lbrace$\ce{Y_{0.67}Mo_{0.33}}$\rbrace$, and {$\lbrace$\ce{Sc_{0.67}Mo_{0.33}}$\rbrace$ with a VEC 4 for the substitution of zirconium, hafnium, and titanium. All substitutions are summarized in figure \ref{fig:sub_XRD}(b). The prepared samples are found to be single phase, and no impurity phases were observed in the XRD patterns. The XRD pattern can be indexed with the (110), (200), (211), (220) and (310) reflections of a lattice with the space group $Im$\={3}$m$, a BCC lattice, with unit cell parameters varying between $a \approx$ 3.29 \AA and 3.33 \AA. At room temperature, elemental Hf, Zr, Ti, Sc, and Y crystallize on an HCP lattice, while Nb, Ta, Cr, and Mo crystallize on a BCC lattice \cite{martensitic,Book_alloys}. The patterns for the substitutions of $\lbrace$\ce{Sc_{0.33}Mo_{0.67}}$\rbrace$ for Nb, {$\lbrace$\ce{Y_{0.67}Mo_{0.33}}$\rbrace$ for Hf, and {$\lbrace$\ce{Sc_{0.67}Cr_{0.33}}$\rbrace$ for Ti are shown in figure \ref{fig:sub_XRD}(c). The maxima of the peak positions and therefore the lattice parameters of the unit cells remain almost unchanged for all the samples. The diffraction peaks are broad, which can be due to several factors: small diffraction coherence volumes (i.e. poor crystallinity), variations in atomic composition in different places in the alloy, or strain broadening of the peaks due to the deformation needed to prepare the samples for the diffraction study. Previous electron microscopy studies indicated a uniformly random elemental distribution in the pristine HEA alloy superconductor \cite{fvrohr_PNAS} and therefore postulating that the peak widths are primarily due to small coherence volumes, we use the Scherrer formula \cite{Scherrer} to estimate the linear dimension of the coherent diffracting volume (i.e. the particle size), which we find to be as small as approximately 200 \AA \ or 50 unit cells. These extremely small volumes are consistent with expectations for a random-crystalline alloy like the ones at hand. The broadest diffraction peaks are observed for the replacements of the group 5 elements. The measured peak intensities are strongly dependent on the preparation of the thin sheet of the alloys in which the diffraction is performed, which induces preferred orientation, and cannot be quantitatively interpreted without more detailed study. \\ \\
\begin{table}
	\caption{The critical temperatures $T_{\rm{c}}$ (in degrees K)for the isoelectronic substitutions with $\rm \lbrace Sc$-Cr$ \rbrace$, $\rm \lbrace$Y-Mo$\rm \rbrace$, and $\rm \lbrace$Sc-Mo$\rm \rbrace$ mixtures in the pristine HEA supercondutor \ce{[TaNb]_{0.67}(HfZrTi)_{0.33}} with $T_{\rm{c}} \approx 7.6$ K.}
	\label{tab:sub}
	\includegraphics[width=0.65\linewidth]{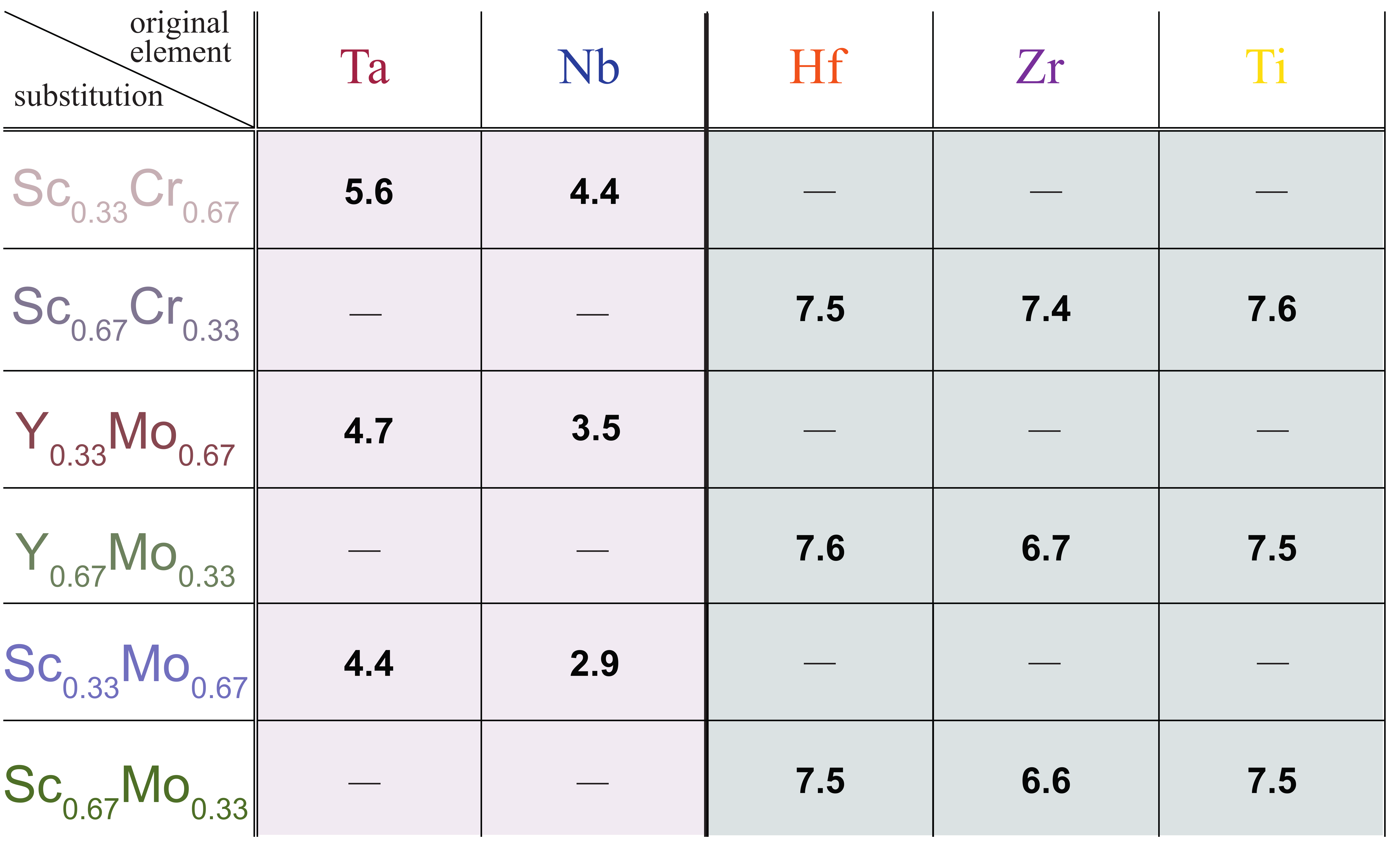}
\end{table}
\textbf{Superconducting Properties of the Substituted HEAs.} In figure \ref{fig:sub_mag}(a)-(c), we show the zero-field cooled (ZFC) magnetization measurements of all the substituted HEAs between $T$ = 2 K and 12 K, in an external magnetic field of $\mu_0 H =$ 2 mT. The pristine HEA \ce{[TaNb]_{0.67}(HfZrTi)_{0.33}} is shown for reference as a black dashed line. All prepared samples are found to be bulk superconductors, with susceptibilities greater than $\chi=$ -1 (i.e. 1/4$\pi$, caused by demagnetization effects, which we correct for better comparability in the figure). The critical temperatures $T_{\rm c}$ were determined as the values at the points where the linearly approximated slopes (dotted line) cross the normal state magnetization, illustrated by an arrow in figure \ref{fig:sub_mag}(b) for the sample \ce{[TaNb]_{0.67}(Hf \lbrace Y_{0.67}Mo_{0.67}\rbrace Ti)_{0.33}}. All transition temperatures are summarized in table \ref{tab:sub}. \\ \\
$T_{\rm c}$ changes significantly upon substitution of the constituent elements. For the replacement of the group IV VEC 4 elements Hf and Ti, the change in $T_{\rm c}$ is negligible for {$\lbrace$\ce{Sc_{0.67}Cr_{0.33}$\rbrace$}, {$\lbrace$\ce{Y_{0.67}Mo_{0.33}$\rbrace$}, and {$\lbrace$\ce{Sc_{0.67}Mo_{0.33}$\rbrace$}. For replacements of Zr, we find a slight lowering of the transition temperature for all three substitutions, with the lowest $T_{\rm c} \approx$ 6.6 K for the sample \ce{[TaNb]_{0.67}(Hf \lbrace Sc_{0.67}Mo_{0.33}\rbrace Ti)_{0.33}}, indicating that the presence of Zr is important for reaching the maximal critical temperature of  $T_{\rm c} \approx$ 7.6 K for the pristine HEA. For the replacement of Nb and Ta by $\lbrace${\ce{Sc_{0.33}Cr_{0.67}}$\rbrace$, $\lbrace${\ce{Y_{0.33}Mo_{0.67}}$\rbrace$, and $\lbrace$\ce{Sc_{0.33}Mo_{0.67}}$\rbrace$, we find a substantial lowering of $T_{\rm c}$. For the replacement of Ta, we find $T_{\rm c} \approx$ 5.6 K, 4.7 K, and 4.4 K, respectively. The lowest $T_{\rm c} \approx$ 4.4 K, 3.9 K, and 2.9 K are observed for the samples where Nb has been substituted by an isoelectronc mixture of elements. This shows that especially the Nb presence in the HEA is important for obtaining an optimal superconducting transition temperature. The critical temperature decreases to $T_{\rm c} \approx$ 2.9 K for \ce{[Ta \lbrace Sc_{0.33}Mo_{0.67}\rbrace]_{0.67}(HfZrTi)_{0.33}}, but the BCC HEA is superconducting above 1.8 K, even when Nb is not present. These findings support the hypothesis that the superconducting properties of these HEAs are not just a compositional average of the properties of the constituent elements, but rather that of a single homogeneous superconducting phase, even for the highly disordered atoms on simple lattices in HEAs. Interestingly, the incorporation of Cr into the HEA has a similar effect on the $T_{\rm c}$ as Mo has, despite the fact that Cr alloys can often be magnetic.
\subsection{Aluminum alloying of the HEA Superconductor}
\begin{figure}
	\centering
	\includegraphics[width=0.48\linewidth]{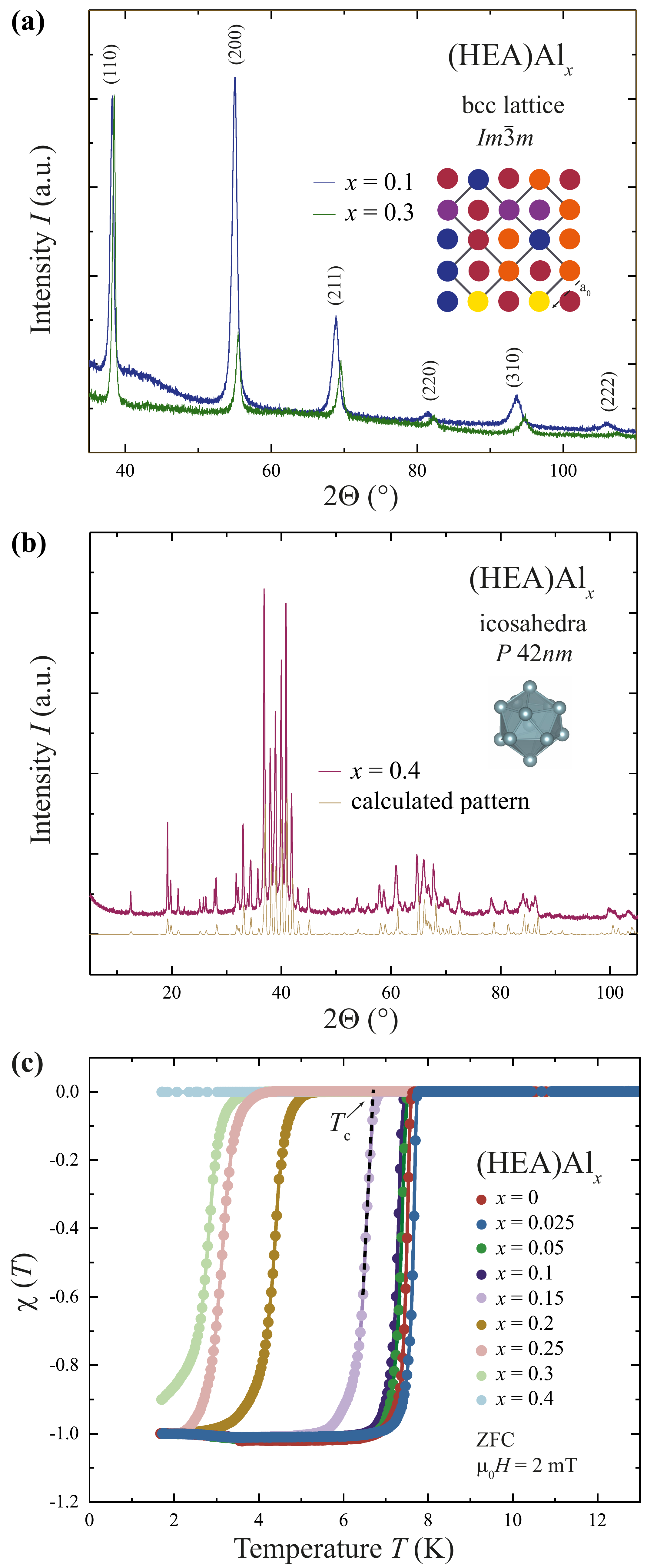}
	\caption{XRD patterns for (HEA)Al$_x$, which (a) arranges on a BBC lattice for $x < 0.4$ (blue and green lines) and (b) crystallizes in the $\beta$-uranium structure type for $x \ge$ 0.4 (red line; the yellow line corresponds to the calculated pattern). (c) Temperature-dependent zero-field-cooled (ZFC)magnetizations for(HEA)Al$_x$ alloys with $x$ = 0, 0.025, 0.05, 0.1, 0.15, 0.2, 0.25, 0.3, 0.4 between $T =$ 1.8 K and 13 K in an external magnetic field of $\mu_0 H =$ 2 mT.}
	\label{fig:dop_XRD}
\end{figure} 
\textbf{The structure of (HEA)Al$_x$.} Aluminum is known to be an excellent alloying element, and it is essential for a wide variety of technologically relevant alloys (see, e.g., reference \cite{aluminum}). Here, we used aluminum, VEC 3, as an addition to investigate the electron count dependence of the superconductivity of the pristine HEA superconductor \ce{[TaNb]_{0.67}(HfZrTi)_{0.33}}. We prepared the samples (HEA)Al$_x$ with $x =$ 0.025, 0.5, 0.1, 0.15, 0.2, 0.25, 0.3, and 0.4. We find that all the alloys crystallize in a BCC lattice up to $x$ = 0.3. The diffraction patterns for $x =$ 0.1 and 0.3 are shown in figure \ref{fig:dop_XRD}(a). The peaks of all samples are broad. However, while the reflection positions shift slightly to higher angles for increasing Al content, they also sharpen. The shifting correlates to the smaller atomic size of aluminum than the average atomic sizes of the components of the HEA. Therefore, when we take the maxima of the reflections, the cell parameters of the BCC lattice decrease linearly from  $a \approx$ 3.34 \AA \ to 3.29 \AA \ within the series. The sharpening of the peaks is due to an increasing crystallinity of the HEA. The crystallinity (followed through the decrease in peak widths) increases until an aluminum content of $x$ = 0.4 is reached. At this point, the high-entropy stabilization of a simple BCC lattice breaks down. Rather, the $x$ = 0.4 material crystallizes in the $\beta$-uranium structure type, with the tetragonal space group \textit{P}4$_2$/\textit{mnm}, and the cell parameters $a \approx$ 9.96 \AA \ and $c \approx$ 5.20 \AA. We show the corresponding XRD pattern in figure \ref{fig:dop_XRD}(b); despite the large number of elements present (seven), the XRD pattern is almost free of any impurity phases, and the BCC lattice has completely disappeared. This material can be considered as a well crystallized alloy, resulting in sharp reflections in the XRD pattern. This phase is commonly also referred to as the $\sigma$ phase. Its body centred tetragonal unit cell contains 30 atoms distributed among five non-equivalent crystallographic sites. The formation of this phase in the 7 element HEA alloy is consistent with the Ta-Al and Nb-Al binary phase diagrams, where the formation of this structure is observed starting at \textit{M}Al$_x$ (\textit{M} = Ta, Nb) $x > 0.36$ \cite{TaAl,NbAl}. We postulate that the structural transition is driven by the formation of icosahedral arrangements of the atoms, a structural motif that is known to be especially stable for aluminum alloys and compounds.  \\ \\
\begin{figure}
	\centering
	\includegraphics[width=0.6\linewidth]{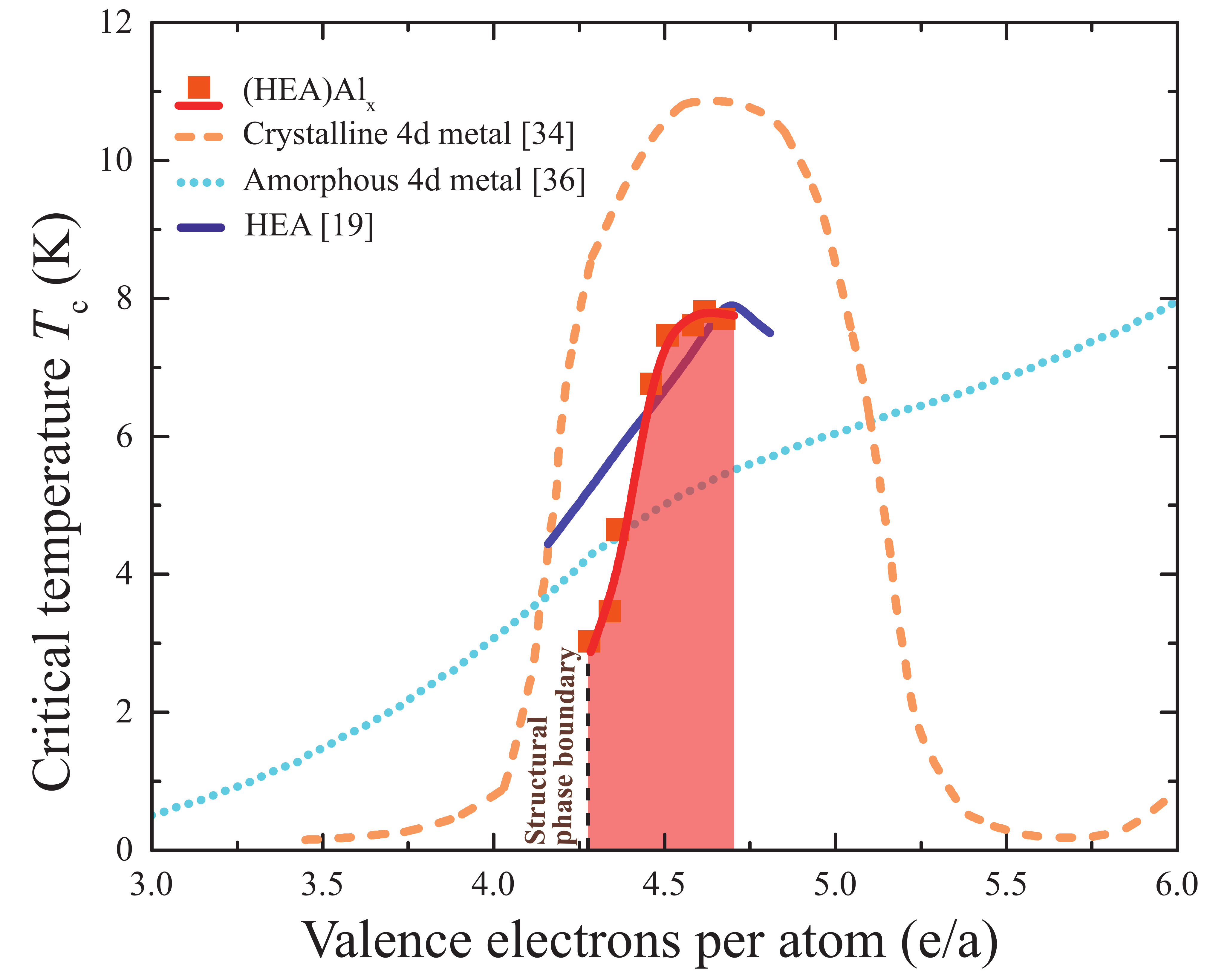}
	\caption{Critical temperatures $T_{\rm c}$ as a function of valence electron count per atom e/a for (HEA)Al$_x$ (red points, red solid line is a trend line). For comparison the observed trend lines of the transition metals and their alloys in the crystalline form (tangerine dashed line) \cite{Matthias,Simon}, of the critical temperatures of the HEA solid solution \ce{[TaNb]_{1-\textit{x}}(ZrHfTi)_{\textit{x}}} (blue solid line) \cite{fvrohr_PNAS}, and as amorphous vapor-deposited films (azzurro dotted line) \cite{amorph1,amorph2} are depicted.}
	\label{fig:dop_mag}
\end{figure}
\textbf{Superconducting Properties and Phase Diagram of (HEA)Al$_x$.} In figure \ref{fig:dop_mag}(a), we show the ZFC magnetization characterizations of all the aluminum alloyed HEAs between $T$ = 2 K and 12 K, in an applied magnetic field of $\mu_0 H =$ 2 mT. All samples with $x <$ 0.4, are found to be bulk superconductors with sharp transitions to the superconducting state. All of the superconducting materials, with the exception of $x$ = 0.3 have measured susceptibilities that are larger than $\chi=$ -1 due to demagnetization effects, and are therefore scaled to the $\chi=$ -1 value. The critical temperatures $T_{\rm c}$ were determined as the values at the points where the linearly approximated slopes in the superconducting transitions (dotted line) intercept the normal state magnetization, illustrated by an arrow in figure \ref{fig:dop_mag}(a) for $x$ = 0.15. We find the transition temperatures to decrease with increasing Al content within the (HEA)Al$_x$ series. For the sample $x = 0.4$, which crystallizes in the $\beta$-uranium structure, no transition to a superconducting state can be observed above $T =$ 1.8 K. Whether the implied complete disappearance of superconductivity at this structural phase boundary actually occurs or whether it is merely suppressed to below 1.8 K will require further study. If the superconductivity is indeed fully suppressed at the structural phase boundary, as is known for many other materials (see, e.g. references \cite{BiO_SC,WO3,Ram,Daigo,Bendele, Weyeneth}), then the superconducting HEA system studied here offers a good opportunity to investigate this suppression on the fundamental BCC crystal lattice. \\ \\
The determined critical temperatures $T_{\rm c}$ for the HEAs are plotted in figure \ref{fig:dop_mag}(b) as a function of the valence electron (VE) to atom ratio $e/a$ (red squares, red solid line is a trend line). For comparison, the observed trend lines for the critical temperatures of the HEA solid solution \ce{[TaNb]_{1-\textit{x}}(ZrHfTi)_{\textit{x}}} \cite{fvrohr_PNAS} (blue solid line), the transition metals and their alloys in crystalline form \cite{Matthias,Simon} (tangerine dashed line) and amorphous vapor-deposited films \cite{amorph1,amorph2} (azzurro dotted line) are also shown. The trend for transition metals, known as the \textit{Matthias rule}, links the critical temperature maxima with the VEC in simple binary transition-metal phases. The trend line for amorphous superconductors is from the work of Collver and Hammond and coworkers, who studied the $T_{\rm c}$ of vapor-cryodeposited films of simple transition-metal alloys and came to the conclusion that  $T_{\rm c}$ vs. e/a ratio no longer followed the characteristic behavior of the \textit{Matthias rule} for crystalline binary alloys: in amorphous transition-metal thin films, the critical temperatures $T_{\rm c}$ were found to increase with increasing e/a, in a monotonic almost structureless way. The maximum is reached at a much higher valence electron to atom ratio of e/a = 6.4. These two curves, the \textit{Matthias rule}, and the amorphous critical temperatures $T_{\rm c}$ after Collver and Hammond are the established standards to which other superconductors may be compared. We find that the electron count dependence of the critical temperatures in (HEA)Al$_x$ falls between the two benchmark lines, but that it rather follows the \textit{Matthias rule}. Especially, compared to the trend line of the \ce{[TaNb]_{1-\textit{x}}(ZrHfTi)_{\textit{x}}} solid solution, it becomes evident that the incorporation of aluminum causes a more crystalline-like critical temperature dependence. This is also in good agreement with the tendency toward more crystallinity observed in the XRD patterns, and the transition to an ordered intermetallic compound. These findings are in contrast to those on the \ce{[TaNb]_{1-\textit{x}}(ZrHfTi)_{\textit{x}}} solid solution, where an almost amorphous-like behavior was observed for the whole solid solution range. 
\section{Summary and Conclusion}
We have investigated the HEA superconductor \ce{[TaNb]_{1-\textit{x}}(ZrHfTi)_{\textit{x}}} by isoelectronic substitutions, and aluminum alloying. We have synthesized, by arc-melting, samples of the HEA superconductor \ce{[TaNb]_{0.67}(ZrHfTi)_{0.33}}, substituted with isoelectronic mixtures of $\rm \lbrace Sc$-Cr$ \rbrace$, $\rm \lbrace$Y-Mo$\rm \rbrace$, and $\rm \lbrace$Sc-Mo$\rm \rbrace$. We also alloyed the HEA superconductor with aluminum according. For the isoelectronic substitutions, we found that all materials crystallize on a BCC lattice ($Im\bar{3}m$), with the cell parameters remaining relatively unchanged. All prepared samples of this substitutional series are found to be bulk superconductors, with strongly changing critical temperatures, depending on the chemical composition of the HEA. We conclude that the superconducting transition temperatures strongly depend on the elemental makeup of the alloy, despite the large degree of disorder and the random mixing of the elements. This is most pronounced for Nb and also Ta, where in the HEA \ce{[Ta  \lbrace Sc_{0.33}Mo_{0.67}\rbrace]_{0.67}(HfZrTi)_{0.33}} the critical temperature $T_{\rm c}$ decreases more than 60 \% by the isoelectronic replacement. For replacements of the group IV elements, the changes in critical temperature are much weaker. While the $T_{\rm c}$ stays virtually unchanged for the replacement of Hf and Ti, we observe a decrease of up to $\Delta T \approx 1$ K for the replacement of Zr. Overall the effect of introduced disorder by increasing the amount of constituent atoms does not lead to a loss of the superconductivity or a large decrease of the critical temperature $T_{\rm c}$. For the alloying of aluminium into the near optimal VEC \ce{[TaNb]_{0.67}(ZrHfTi)_{0.33}} superconductor, as in (HEA)Al$_x$, we find single phase XRD patterns that can be indexed with a BCC lattice ($Im\bar{3}m$) up to $x =$ 0.3. With increasing Al content, the cell parameter decreases from $a \approx$ 3.34 \AA \ to 3.29 \AA, while at the same time, the broad diffraction peaks of the HEA sharpen. This reflects an increasing crystallinity of the alloys. We found that for $x =$ 0.4 the high-entropy stabilization of a simple BCC lattice breaks down. This material crystallizes in the $\beta$-uranium structure type, with the tetragonal space group \textit{P}4$_2$/\textit{mnm}. Superconductivity, if present at all, is suppressed to temperatures below 1.8 K. We find that the electron count dependence of the superconducting critical temperatures in (HEA)Al$_x$ (up to $x <$ 0.4) falls between the benchmark lines for amorphous and crystalline transition-metal superconductors, following a more crystalline-like critical temperature dependence on VEC than the \ce{[TaNb]_{1-\textit{x}}(ZrHfTi)_{\textit{x}}} supeconductor, where an almost amorphous-like behavior was observed for the whole solid solution.  This is consistent with the powder diffraction characterization of these materials, which shows improved crystallinity of the BCC HEA when Al is present. Our results suggest that HEAs are an interesting model system for the investigation of structure-property relations in "simple" intermetallic superconductors.
\section*{Acknowledgments}
This work was supported by the Gordon and Betty Moore Foundation, EPiQS initiative, Grant GBMF-4412.


\begin{thebibliography}{99}
\bibitem{note} For better readability of the chemical formula, all elements with a valence electron count (VEC) of 5 VE are written in square brackets, while elements with a of VEC 4 are written in round brackets throughout the manuscript. Further, though the compositions of HEAs are generally presented on the basis of 100 atoms per formula unit, here we employ formulas that yield a total of 1 atom per formula unit; this difference resulted in ambiguity in our earlier publication \cite{fvrohr_PNAS}, where both nomenclatures were employed, particularly concerning the magnitudes of the specific heat jumps at $T_{\rm{c}}$.
\end{thebibliography}
 \end{document}